\documentclass[lineno,authoryear]{FLO_v1}%

\usepackage{graphicx}
\usepackage{upgreek}
\usepackage{multicol,multirow}
\usepackage{amsmath,amssymb,amsfonts}
\usepackage{mathrsfs}
\usepackage{amsthm}
\usepackage[figuresright]{rotating}
\usepackage{appendix}
\usepackage[authoryear]{natbib}
\usepackage{ifpdf}
\usepackage[T1]{fontenc}
\usepackage{newtxtext}
\usepackage{newtxmath}
\usepackage{textcomp}
\usepackage{xcolor}
\usepackage{epstopdf, epsfig}
\usepackage{subcaption}
\usepackage{graphbox}
\usepackage{csquotes} 
\usepackage{multirow}
\usepackage{array}
\usepackage{float}
\usepackage{svg}
\usepackage{booktabs}
\usepackage{siunitx}
\usepackage{placeins}
\usepackage[skip=8pt]{caption}

\usepackage[colorlinks,allcolors=blue]{hyperref}
\definecolor{jourcolor}{cmyk}{1,0.57,0.01,0.38}
\definecolor{lightred}{rgb}{1,0.35,0.35}
\hypersetup{
    colorlinks,%
    citecolor=jourcolor,%
    filecolor=jourcolor,%
    linkcolor=jourcolor,%
    urlcolor=jourcolor
}

\makeatletter
\renewcommand{\NAT@open}{\textcolor{jourcolor}{(}}   
\renewcommand{\NAT@close}{\textcolor{jourcolor}{)}}  
\makeatother

\theoremstyle{definition}

\articletype{RESEARCH ARTICLE}

\DOI{Preprint submitted to Flow}

\citearticle{Anand, A., Nikolaidou, L., Poelma, C. \& Laskari, A. }

\begin{document}

\title[Turbulent boundary layer development over an air cavity]{Turbulent boundary layer development over an air cavity}

\author[A. Anand, L. Nikolaidou, C. Poelma and A. Laskari]{Abhirath Anand$^{1,\ast}$, Lina Nikolaidou$^{1}$, Christian Poelma$^{1}$  and Angeliki Laskari$^{1}$}

\address[1]{Multiphase Systems, Department of Process and Energy, Delft University of Technology, The Netherlands }

\corres{*}{Corresponding author. E-mail:
\emaillink{a.anand-1@tudelft.nl}}


\date{\textbf{Received:} XX XXXX; \textbf{Revised:} XX XX XXXX; \textbf{Accepted:} XX XX XXX}

\abstract{The turbulent boundary layer (TBL) development over an air cavity is experimentally studied using planar particle image velocimetry. The present flow, representative of those typically encountered in ship air lubrication, resembles the geometrical characteristics of flows over solid bumps studied in literature. However, unlike solid bumps, the cavity has a variable geometry inherent to its dynamic nature. An identification technique based on thresholding of correlation values from particle image correlations is employed to detect the cavity. The TBL does not separate at the leeward side of the cavity owing to a high boundary layer thickness to maximum cavity thickness ratio ($\delta/t_{max}= 12$). As a consequence of the cavity geometry, the TBL is subjected to alternating streamwise pressure gradients: from an adverse pressure gradient (APG) to a favourable pressure gradient and back to an APG. The mean streamwise velocity and turbulence stresses over the cavity show that the streamwise pressure gradients and air injection are the dominant perturbations to the flow, with streamline curvature concluded to be marginal. Two-point correlations of the wall-normal velocity reveal an increased coherent extent over the cavity and a local anisotropy in regions under an APG, distinct from traditional APG TBLs, suggesting possible history effects. }

\maketitle

\begin{boxtext}

\textbf{\mathversion{bold}Impact Statement}

With the maritime industry's focus on reduced emissions and sustainability, drag reduction in ships is receiving growing attention. Air lubrication, a frictional drag reduction method, involves injecting air beneath the ship's hull to minimize its contact with the surrounding, often turbulent flow. The morphology and dynamics of the air cavity formed, aspects crucial to air lubrication, are intrinsically linked to the incoming flow conditions. We experimentally investigate one aspect of this two-way coupling: the response of a turbulent boundary layer (TBL) to an air cavity formed under a specific incoming flow condition. Using single and two-point statistics, we characterize the TBL development over the air cavity and examine the influence of different cavity regions on the TBL. Results and conclusions from this study are expected to not only provide insights regarding air cavity formation, but also establish a baseline reference for future studies examining variable inflow conditions and external perturbations. 

\end{boxtext}

\section{Introduction}
\label{sec: intro} 

Turbulent flows are ubiquitous in nature and many engineering applications, and are often subject to external perturbations that can emanate from the free-stream or the geometry over which they develop in case of wall-bounded flows. For instance in nature, the topographical features of the earth in the form of ``surface roughness'' can perturb the often turbulent, atmospheric boundary layer. Flows encountered in engineering applications such as those over curved geometries (for example airfoils) or in turbomachinery, can be subject to a range of pressure gradients and curvatures that can have significant influence on the flow and hence, the performance of the system.

Within this context, the creation of an air cavity in the path of an incoming turbulent boundary layer (TBL) is an example typically encountered in air lubrication or air layer drag reduction (ALDR) employed in ships \citep{elbing2008, makiharju2012}. Briefly, the method of ALDR involves the injection  of air (above a critical flow rate) below the ship hull which develops into a layer or a cavity, separating the surrounding liquid and the solid surface and thereby significantly reducing the skin-friction drag. The air cavity thickness, its length and stability, important parameters in the technique of ALDR, have been reported to be dependent on the incoming flow conditions \citep{zver2014,nikolaidou2024}. In addition to the incoming flow conditions and the air flow rate employed, the morphology and stability of the air cavity formed is dependent on the way the air is introduced below the ship hull. Common methods include employing a cavitator (similar to a backward-facing step) \citep{zver2014} or a slot-type injector \citep{elbing2008,nikolaidou2024}. In this study, we report experimental findings on a typical case within the method of slot-injection: the behaviour of an incoming TBL encountering a slot-injected air cavity formed at a specific incoming flow condition and air flow rate.

 The flow over the air cavity encountered here is in some ways geometrically similar to solid bump flows studied in literature (for example: \citet{baskaran1987,webster1996}). The incoming TBL is perturbed by the obstruction created by an air cavity in its path, and this creates a local change in boundary condition. The change in the local boundary condition from a no-slip at the wall to a free-slip boundary, the presence of a relatively variable obstacle geometry inherent to the dynamic nature of the air cavity, and the presence of multiple pressure gradients and curvatures, significantly increases the complexity of the current study relative to solid bump flows.

 In an experimental study of a TBL moving over a two-dimensional curved hill by \citet{baskaran1987}, the TBL was subjected to multiple streamwise pressure gradients brought about by changes in surface curvature. A complex pressure gradient pattern alternating in sign, from an adverse pressure gradient (APG) to a favourable pressure gradient (FPG) and back to an APG was imposed by the solid bump. The TBL was found to separate at the leeward side of the bump and this was attributed to a low incoming boundary layer thickness to bump height ratio ($\delta/h = 0.25$). An experimental study with a similar flow geometry of a TBL flow over a solid bump with a $\delta/h = 1.5$ reported {\em{no}} separation of the TBL at the leeward side \citep{webster1996}, highlighting $\delta$ and $h$ as important length scales in these type of flow geometries. The response of the TBL to a sequence of pressure gradients in terms of the mean velocity or the turbulence stresses has been found to be qualitatively and quantitatively different compared to its response towards individually and continuously applied pressure gradients. For instance, the appearance of inflectional points in the Reynolds stress profiles known as ``knee'' points, common footprints of a developing internal layer, has been reported in TBLs subjected to a sequence of alternating pressure gradients \citep{webster1996,balin2021,aadhy2023}. The sudden change in boundary conditions, and hence the wall shear stress, has been reported to be responsible for the development of these layers \citep{smits1985}. They are known to carry the disturbances induced by streamwise pressure gradients \citep{balin2021} and/or discontinuities in surface curvature \citep{baskaran1987,webster1996} from the inner to the outer region, and play an important role in the characteristics of the recovering TBL.

The effects of different perturbations such as pressure gradients, surface curvatures, wall roughness, etc., on a TBL have been extensively reviewed by \citet{smits1985}. The response and recovery of TBLs were observed to depend on the strength and the distance over which the perturbations were applied. The inner and outer regions of the TBL were found to behave and respond differently based on the type and distributions of the perturbations imposed. For example, FPGs and APGs have opposing effects on a TBL. TBLs subjected to FPGs accelerate and as a result experience a decay in turbulence stresses, which is most pronounced in the outer region. This is a consequence of a reduction in turbulent kinetic energy (TKE) production due to weakened outer-region structures \citep{bourassa2009,harun2013,volino2020}. For sufficiently high accelerations, commonly characterised by the acceleration parameter $K = (\nu$/$U_{e}^2$)($dU_{e}$/$dx$), where $\nu$ is the kinematic viscosity and $U_e$ is the local external velocity, TBLs can reach a quasi-laminar state or even undergo relaminarisation \citep{narasimha1979}. In addition, the wake region can seize to exist and the logarithmic region loses universality \citep{bourassa2009}. On the other hand, TBLs subjected to APGs decelerate and undergo an amplification in turbulence stresses, most pronounced in the outer region, owing to increased TKE production by large-scale structures \citep{monty2011,harun2013}. A dip in the mean velocity profile below the log-law and a shortening of the logarithmic region, along with an increased wake have also been reported \citep{nagano1988,monty2011}. It is important to point out that experimentally, the application of FPGs or APGs or a combination of both to TBLs can be performed in a variety of ways, such as using variable height tunnels, ramps, or curvatures, each approach likely to have a different influence on the developing TBL.

The imposition of traditional APGs or FPGs on TBLs, where the pressure gradient is continuously applied, has been reported to affect the spatial coherence of turbulent structures. The streamwise extent of coherence has been observed to reduce in APG TBLs compared to zero pressure gradient (ZPG) TBLs, whereas the wall-normal one was found to be virtually unaffected \citep{krogstad1995}. TBLs subjected to a FPG have been observed to exhibit an increased extent of coherence in both the streamwise and wall-normal directions \citep{volino2020}. In more complex pressure gradient impositions, such as in alternating pressure gradients imposed by a solid bump, the influence of upstream conditions or history effects become important \citep{baskaran1987,webster1996,aadhy2023}. These upstream effects are expected to influence the spatial coherence of turbulent structures in a different way compared to traditional pressure gradient TBLs due to the fact that FPGs and APGs evoke opposing responses from turbulent structures. However, it is not known how such sequential pressure gradients affect the spatial coherence of turbulent structures compared to traditional FPG/APG TBLs and ZPG TBLs.

Given the similarity in flow geometry of solid bumps studied in literature to the air cavity in the present study, the imposition of multiple pressure gradients and curvatures in addition to a free-slip boundary condition is expected to influence the incoming TBL. The latter has been shown to be important in the resulting morphology and dynamics of the air cavity, crucial aspects in the application of ALDR. However, studies on the behaviour of a TBL in the presence of an air cavity remain scarce. In the present study, we analyse one aspect of the two-way coupling, that is the response of a TBL in the presence of an air cavity. Experiments using planar particle image velocimetry (PIV) are carried out to provide qualitative and quantitative insights into the TBL response and development. 
 Details of the experimental technique including the parameters of data acquisition and processing are given in \hyperref[sec:expt-setup]{section~2}. Characteristics of the baseline TBL, the identification of the air cavity, and the statistics of the development of the TBL are presented and discussed in \hyperref[sec:results]{section~3}. Conclusions and potential for future work are presented in \hyperref[sec:conclusion]{section~4}.

\section{Experimental setup}
\label{sec:expt-setup}

Experiments were carried out in the water tunnel of the Laboratory for Aero and Hydrodynamics at TU Delft. The test section of the water tunnel has a length of $5\;\mathrm{m}$ and a cross-sectional area of $0.6\times0.6\;\mathrm{m}^2$. The free-surface of the water tunnel is covered by two plates, each $2.485\;\mathrm{m}$ in length, beneath which a liquid boundary layer develops. In order to ensure transition to a TBL, a zigzag strip of thickness $0.5\;\mathrm{mm}$ is used at the leading edge of the first plate. The characteristics of the incoming turbulent boundary layer at the measurement location are summarised in \hyperref[tab:tbl_baseline]{table~1}. The boundary layer thickness in the current study is estimated using turbulent quantities (see \hyperref[sec: turb_stats]{section~3.3} for further discussion). The thickness estimated using $\delta_{99}$ was found to be comparable to estimates by \citet{nikolaidou2024} at the same location measured in a separate experimental campaign. In what follows, $x$ and $y$ will denote the streamwise and wall-normal coordinates, respectively, with $\overline{u}$ ($u'$) and $\overline{v}(v')$ the corresponding mean (fluctuating) velocities.

\begin{figure}
  \centering 
  \includegraphics[width=0.91\textwidth]{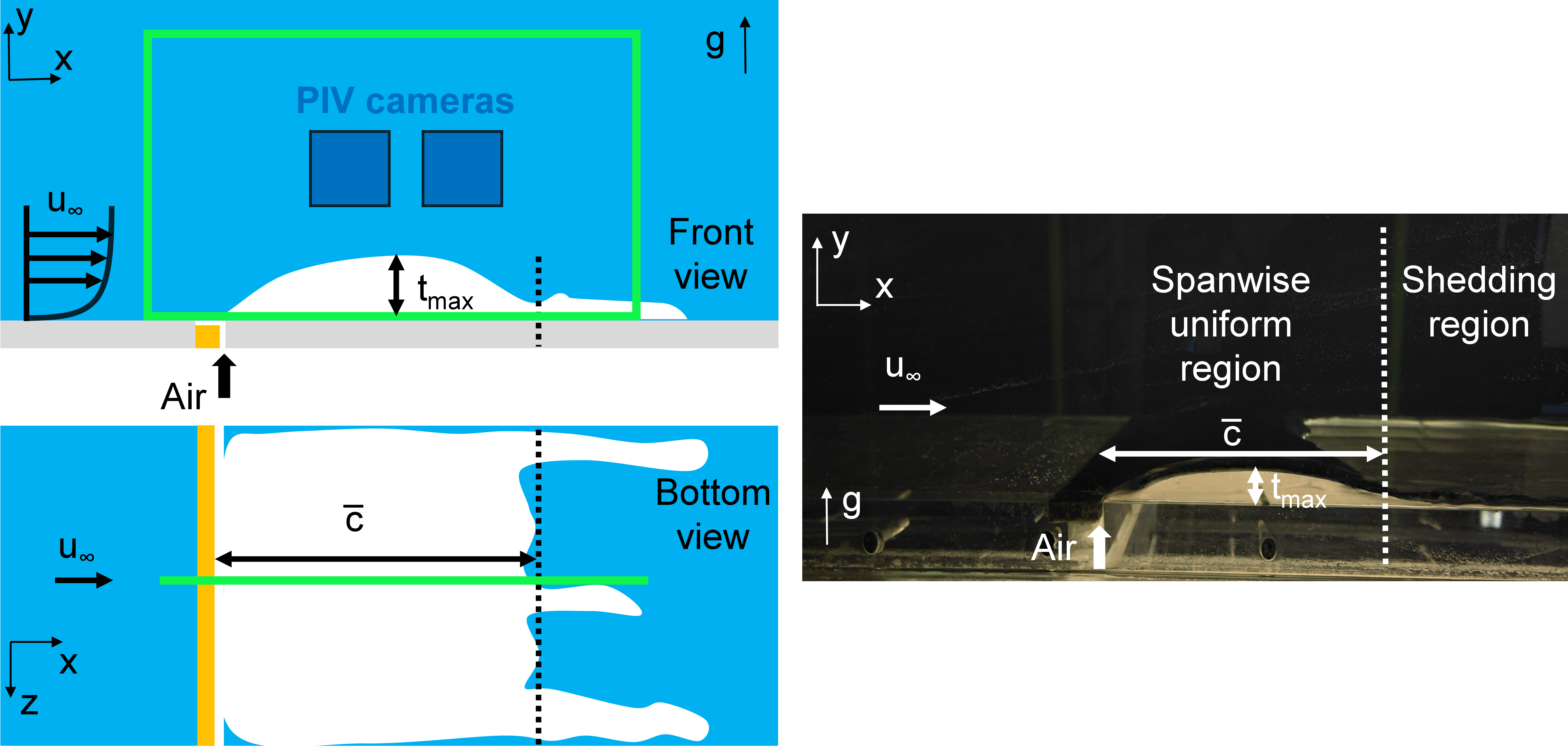}% Images in 100% size
  \caption{Schematic of the experimental setup (left). The laser illuminated field of view is shown in green. The cavity (in white) shown upside down in line with the flow geometry encountered in solid bump flows. The cavity's main geometrical characteristics ($t_{max}$ and $\overline{c}$) are schematically shown in front and bottom views. Two distinct regions can be identified (separated by a dotted black line): an approximately stable spanwise-uniform region upstream and an unstable shedding region downstream. An actual image of the cavity (front view) is also shown on the right. Flow is from left to right. Gravity (g) is from bottom to top. }
  \label{fig:expt_setup}
\end{figure}

The air injection is carried out by a slot-type injector, located at a streamwise distance of $18\delta$ from the leading edge of the first plate, with a gap of $4\;\mathrm{mm}$ in $x$ and spanning the entire width of the plate in $z$. Air is injected perpendicularly into the flow (at a flow rate of $62\;\mathrm{L/min}$) and, as a result of the inertia carried by the liquid, it bends, reaching the surface of the plate forming an air cavity (see \hyperref[fig:expt_setup]{figure~1}; note that the images are rotated to have the cavity at the bottom to match prior bump studies). The morphology of the air cavity formed is such that two distinct regions can be identified from visual observations: an almost stable region upstream (left of the dotted black line in \hyperref[fig:expt_setup]{figure~1}) that is uniform in the span and a shedding region downstream (right of the dotted black line in \hyperref[fig:expt_setup]{figure~1}) varying in the span and comprising thin sheets of air (see also supplementary movie $1$). Given the higher thickness of the spanwise-uniform region compared to that of the  shedding region, the former is expected to impose a much larger perturbation to the incoming flow and thus it will be our focus in this study. With the term cavity, we will then refer to the upstream, spanwise-uniform region of the air cavity in what follows.

The behaviour of the turbulent boundary layer in the presence of the air cavity is studied using planar PIV in a streamwise--wall-normal plane. The flow is seeded with $15\; \mathrm{\mu m}$ hollow glass spheres that are illuminated by a Litron dual-cavity Nd-YAG laser. 
Two LaVision Imager sCMOS CLHS cameras ($5\;\mathrm{Mpix}$, $16$-bit) are used to capture the particle images, placed in a side-by-side configuration, each of them fitted with Nikkor objectives of focal length $105\;\mathrm{mm}$. The resulting magnification is approximately $11.8\;\mathrm{pix}/\mathrm{mm}$ and the combined field of view (FOV) extents $2.87\delta \times 1.35\delta$ in $x$ and $y$, respectively. A total of $2500$ uncorrelated image pairs are acquired at a frequency of $3\;\mathrm{Hz}$ and subsequently processed using Davis $10.1.2$, following a multi-pass approach, with a final window size of $48\times48\; \mathrm{pix}$ and an overlap of $50\;\%$. The resulting final vector spacing is $2\;\mathrm{mm}$ ($\Delta l^+ = 52.6$ in viscous units).

\section{Results and Discussion}
\label{sec:results}

\subsection{Baseline turbulent boundary layer characteristics}

\begin{figure}
\centering
\begin{subfigure}[t]{0.49\textwidth}

\includegraphics[width =1.04\textwidth]{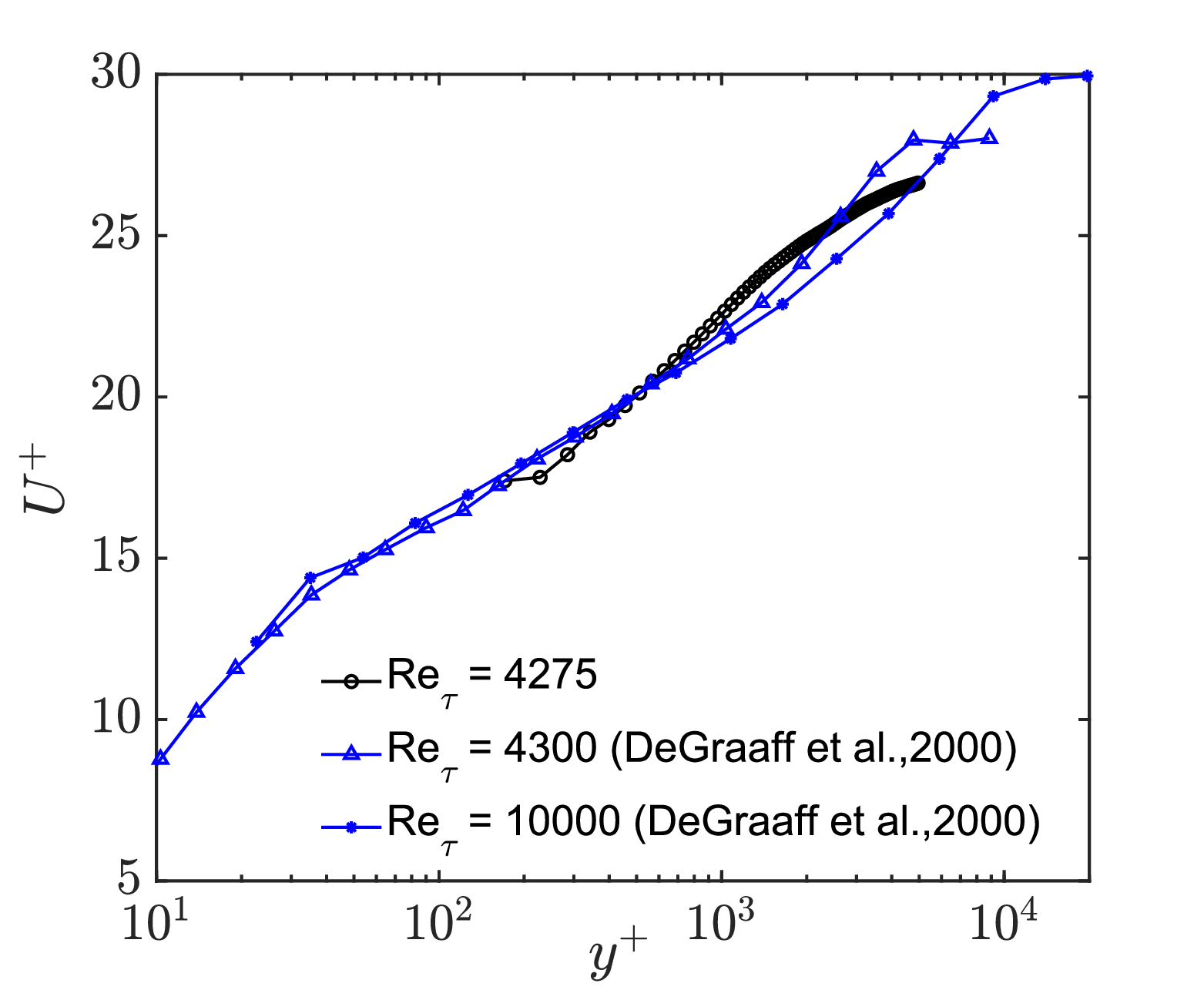}
\caption{}
\label{fig:baseline_mean}
\end{subfigure}
\medskip
\begin{subfigure}[t]{0.49\textwidth}

\includegraphics[width =1.04\textwidth]{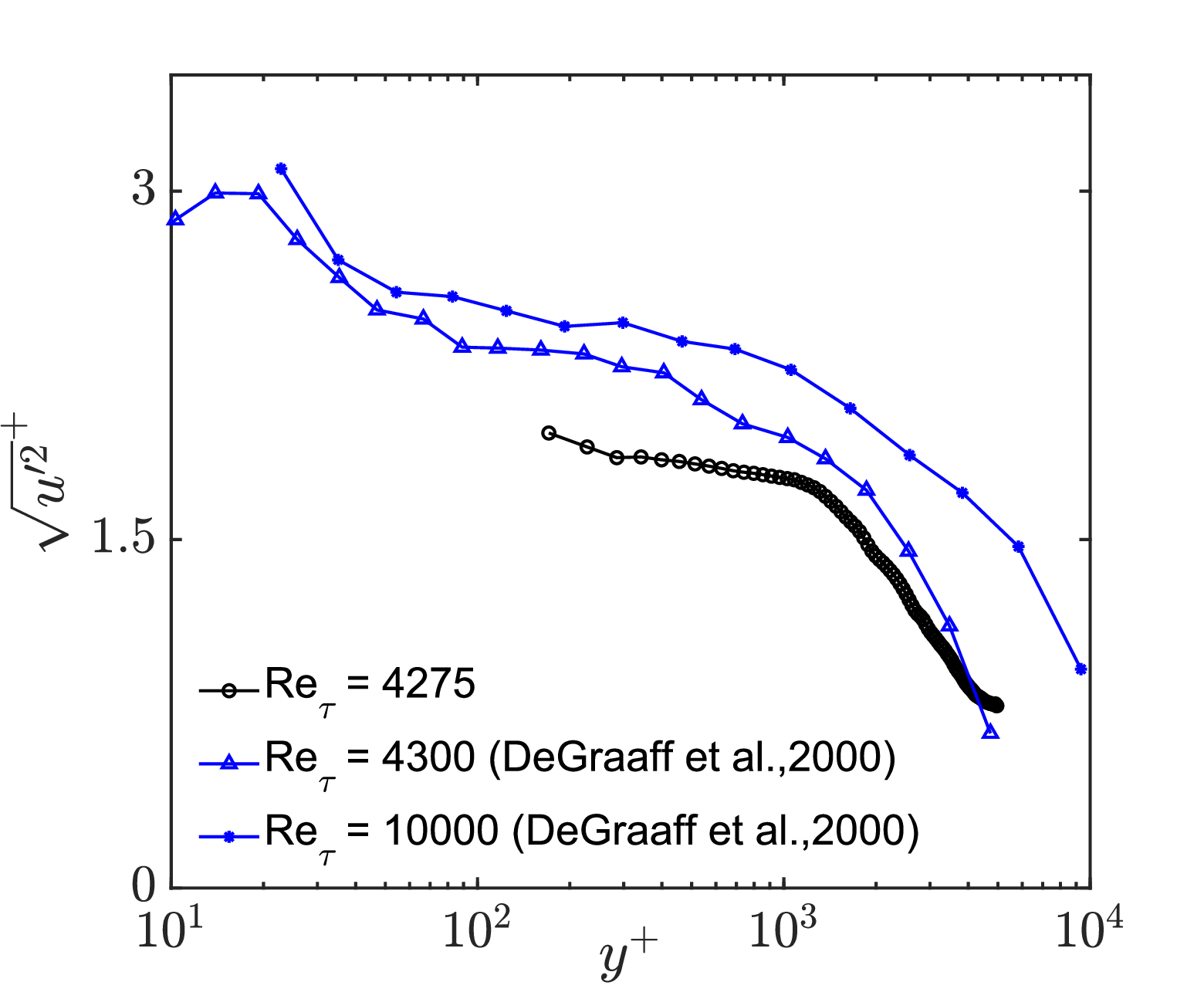}
\caption{}
\label{fig:baseline_stress}
\end{subfigure}
\caption{Inner normalised (a) mean streamwise velocity and (b) streamwise velocity intensities for the baseline TBL. Also shown is reference data using LDA by \citet{degraaff2000}.}
\label{fig:baseline_tbl}
\end{figure}

The incoming liquid turbulent boundary layer developing over the wall is first characterised. This is done in the absence of air injection. The presence of the latter was found to impose an upstream APG on the incoming TBL \citep{anand2021}. Therefore, to avoid any complications while comparing the behaviour of the TBL over the cavity with respect to the baseline case (over the wall), we characterise the incoming TBL without the cavity. 
The inner-normalised mean streamwise velocity and root-mean-square (rms) streamwise turbulent velocity fluctuations of the baseline TBL are shown in \hyperref[fig:baseline_tbl]{figure~2}. In comparison with Laser Doppler Anemometry (LDA) measurements of \citet{degraaff2000} at comparable Reynolds number, the logarithmic behaviour in the present incoming TBL is not strictly canonical. The absence of a well-formed wake can be explained as a result of a freestream that is more noisy than expected. The underestimation in the rms streamwise velocity fluctuations observed here compared to the LDA measurements can be explained due to the relatively large interrogation areas employed. The nature of the incoming turbulent boundary layer in the present study, although not strictly canonical, is not expected to affect the forthcoming results as the aim is to study the relative changes to the TBL as it develops over the air cavity.

\begin{table}
  %\begin{center}
    \centering

  \begin{tabular}{lccc}
  \toprule
    
      $u_{\infty}$ (m/s)   & $u_{\tau}$ (m/s)   &   $\delta$ (m) & $Re_{\tau}$ \\[3pt]
       ~~0.698   & 0.0263 & ~0.159~ & 4275\\
       \bottomrule
  \end{tabular}
  \caption{Baseline TBL characteristics (without cavity)}
  \label{tab:tbl_baseline}
  %\end{center}
\end{table}

\subsection{Air cavity identification}

 As discussed in \hyperref[sec: intro]{section~1}, in TBL flows over solid bumps, the shape of the bump and in particular its height and length, have been observed to influence the development of the TBL \citep{baskaran1987,webster1996}. Given the similarity in geometry of the spanwise-uniform region of the cavity with solid bumps in literature, the shape of the cavity is then also expected to influence the development of the incoming TBL. Therefore, before studying the behaviour of the TBL over the cavity and comparing the flow development with previous solid bump studies, the time-averaged geometry of the cavity needs to be identified and characterised. To this end, an identification technique based on the thresholding of correlation values obtained after image post-processing was employed.

\begin{figure}
  \centering 
  \includegraphics[width=0.72\textwidth]%
  {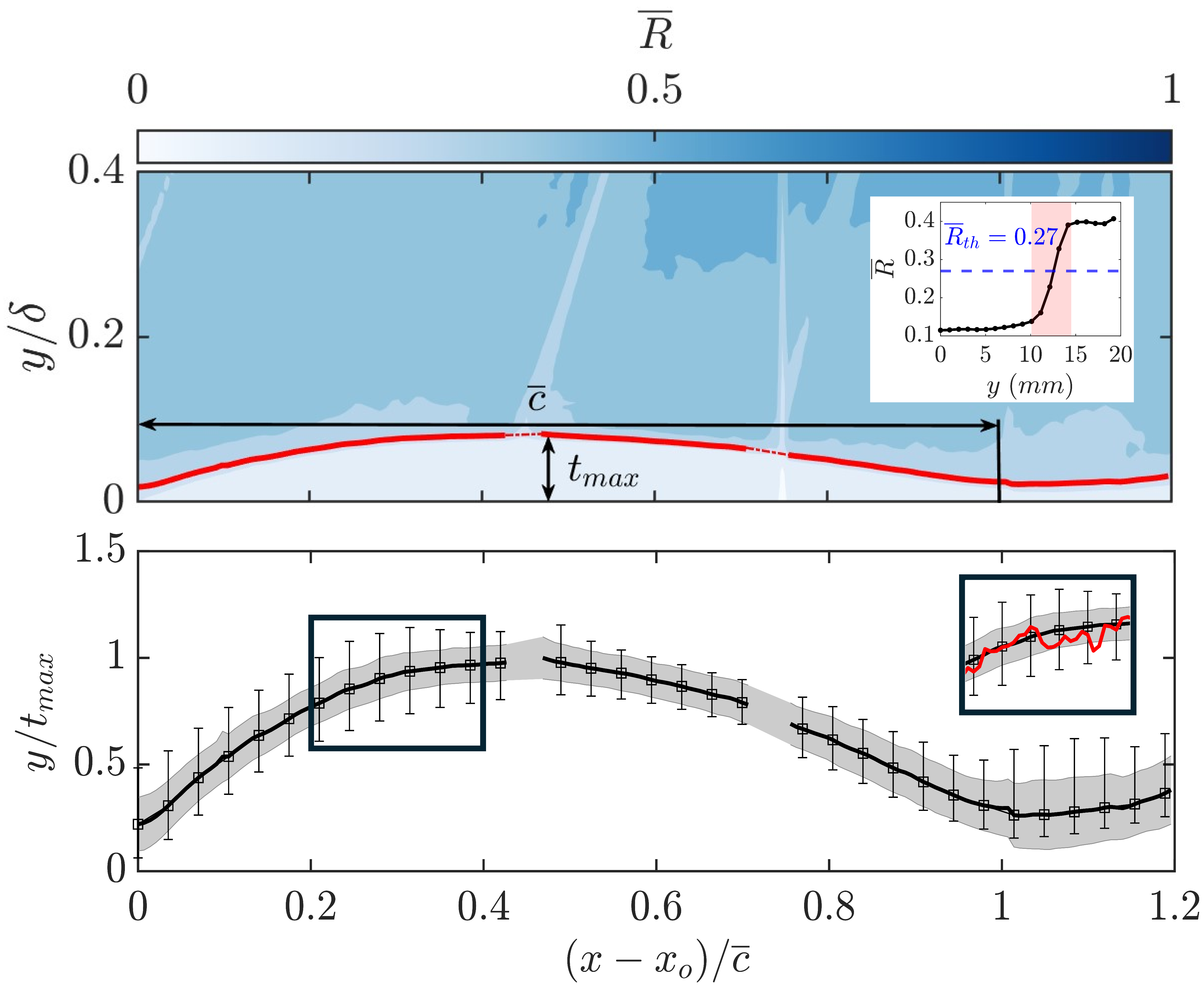}% Images in 100% size
  \caption{Top: Contour of mean correlation map $\overline{R}$ along with the mean cavity identified (in red). Here, $\overline{c}$ is the chord length, $t_{max}$ is the maximum thickness of the mean cavity, and $x_{o}$ is the leading edge of the mean cavity. Inset: Variation of $\overline{R}$ with wall-normal distance at the streamwise location of $t_{max}$ ($\overline{R}_{th}$: correlation threshold chosen to identify the cavity). Bottom:  Mean cavity interface (solid black line) with root-mean-square fluctuations of the cavity thickness along the length of the cavity (shaded grey region). Error-bars depict the uncertainties in the identification technique. Inset: Example of an instantaneous cavity interface detected (in red). See also supplementary movie $1$. }
\label{fig:cavity_detection2}
\end{figure}

The instantaneous correlation values are obtained through a PIV routine following a multi-pass approach, with a final window size of $24\times 24\;\mathrm{pix}$ and an overlap of $50\;\%$ (see \hyperref[fig:cavity_detection2]{figure~3}).  
A smaller final window size (higher spatial resolution) than the one applied to acquire the velocity fields, allowed a more accurate estimate of the air-water interface, and therefore the cavity thickness and length through the correlation values. It was, however, found to be too noisy for velocity estimates; hence, a lower resolution final interrogation window was used there (see \hyperref[sec:expt-setup]{section~2}). 
 A threshold to identify instantaneous geometries of the cavity is obtained using the time-averaged correlation map (\hyperref[fig:cavity_detection2]{figure~3} top). We exploit the fact that correlation values within the cavity should be inherently low compared to the surrounding liquid as there are no tracer particles to correlate with, while patterns within the cavity region (e.g. reflections) also varied significantly between frames and thus did not correlate. This leads to a sharp gradient in correlation between cavity and flow, clearly delineated in the time-averaged map ($\overline{R}$, see inset of \hyperref[fig:cavity_detection2]{figure~3} top).  
 Given the fact that the cavity interface can lie anywhere within this gradient (in the range of $0.15<\overline{R}< 0.40$: see shaded red area in inset of \hyperref[fig:cavity_detection2]{figure~3} top), an average of $\overline{R}$ in this region is identified as a representative threshold ($\overline{R}_{th}=0.27$). 
 This results in an uncertainty (errorbars in \hyperref[fig:cavity_detection2]{figure~3} bottom) in the location of the cavity interface of approximately $\pm2\;\mathrm{mm}$ along its length and dominates any spatial resolution effects. Regardless, the chosen threshold does not affect the analysis in the trends discussed later. This threshold is then applied to every image to detect the instantaneous cavity shape, which is found to match relatively well with visual inspection of the raw images (see supplementary movie $1$). The mean air-water interface is then obtained by averaging over these instantaneous realisations; comparison with the mean directly identified from the time-averaged correlation map \citep{nikolaidou2024}, yields similar results ($<1\;\mathrm{mm}$ difference). The discontinuities observed in the detected interface (\hyperref[fig:cavity_detection2]{figure~3}, e.g. near the $\overline{c}$ label) are due to light path obstructions in the raw particle images. The smoothness of the rest of the field however, allowed interpolation for the mean shape in these locations.

The mean cavity shape reveals a well-defined asymmetric bump-like geometry at the spanwise-uniform part of the cavity upstream (also supported by visual observations). The highly dynamic nature of the shedding region of the cavity downstream, its limited thickness and significantly lower correlation gradients, did not allow for a valid identification of the same. The maximum thickness and chord length of the mean cavity were found to be $t_{max} = 13.2\pm2\;\mathrm{mm}$ and $\overline{c} = 143\pm2\;\mathrm{mm}$, respectively. The ratio of the incoming boundary layer thickness to the maximum thickness of the cavity was found to be $\delta/t_{max} = 12$, which is considerably higher than the ratios encountered in the solid bump studies of \citet{baskaran1987} and \citet{webster1996} ($\delta/h = 0.25$ and $\delta/h = 1.5$, respectively).

Unlike solid bumps for which length and thickness are constant, these parameters vary in space and time for the air cavity studied here. However, when looking at the instantaneous cavity shapes (see supplementary movie $1$) and rms fluctuations of the cavity thickness across its length (grey shaded region around the mean cavity in \hyperref[fig:cavity_detection2]{figure~3} bottom), there are no significant deviations of the cavity from a bump-like geometry. The amplitude of instantaneous variations of the cavity interface detected however, are lower than the estimated experimental uncertainty (see inset of \hyperref[fig:cavity_detection2]{figure~3} bottom), and as such no conclusions can be drawn on it. As detailed analysis of small-scale features (capillary waves or ripples) of the gas-liquid interface lies beyond this study’s scope, we will use the average cavity shape in the analysis that follows. Measurement techniques tailored to interfacial physics would better capture these instantaneous features (for example: \citet{vanMeerkerk2020,gomit2022}).

With the mean cavity geometry defined, the flow behaviour around it can then be analysed. In what follows, a modified coordinate system based on the mean cavity geometry will be employed. Using the {\em{instantaneous}} cavity geometry dimensions did not alter any of the presented results, and thus normalisations based on the mean parameters were chosen for the analysis. The normalised streamwise coordinate is defined as $x' = (x-x_{o})/\overline{c}$, with $x_{o}$ being the leading edge of the cavity. The wall-normal coordinate, $y'$ is normalised with $\delta$, and is defined with its zero-location based on the mean interface height, and maintained normal to the flat plate at the interface positions, the latter employed owing to a relatively large $\delta/t_{max}$ (and minimal streamline curvature) \citep{webster1996}.

\subsection{Turbulence statistics}
\label{sec: turb_stats}

\begin{figure}
  \centering 
  \includegraphics[width=0.81\textwidth]{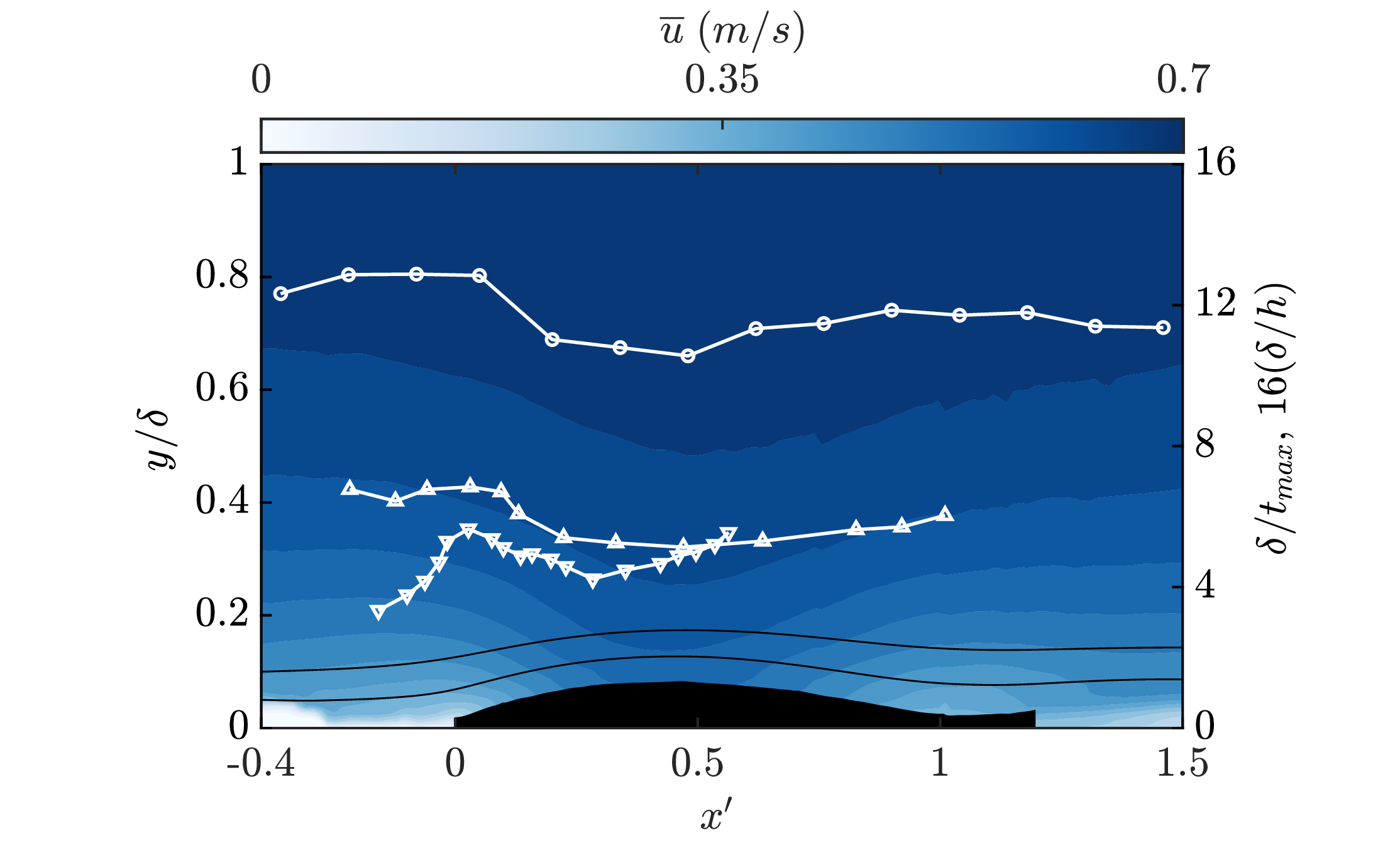}
  %Images in 100% size
  \caption{Mean streamwise velocity contours with some streamlines (solid black lines); $x'=0$ is the leading edge of the cavity where $x' = (x-x_{o})/\overline{c}$. The local $\delta$ from the present study (circles; $\delta/t_{max}=12$) is compared with that over a thicker air cavity (\citet{anand2021}: upward-pointing triangles; $\delta/t_{max}=7$), both normalised by the maximum cavity thickness $t_{max}$, and that over a solid bump (\citet{baskaran1987}: downward-pointing triangles; $\delta/h=0.25$), normalised by the maximum bump height $h$. Flow is from left to right.}
\label{fig:delta_variation}
\end{figure}

 The mean streamwise velocity field clearly illustrates that the incoming TBL experiences an alternating streamwise pressure gradient over the length of the cavity (\hyperref[fig:delta_variation]{figure~4}). The TBL first experiences an APG, as indicated by the moderate elevation of the contour lines at the leading edge of the cavity ($x'=0$). The flow then accelerates around the apex of the cavity, indicating a FPG, while it decelerates past the apex due to an APG, with no indication of incipient separation. 
 To get some further insight on the influence of the cavity on the TBL, the streamwise variation of the local boundary layer thickness $\delta$ is also computed  and compared with the solid bump study by \citet{baskaran1987}. Estimation of $\delta$ in non-equilibrium turbulent boundary layers (for example subjected to pressure gradients) is typically difficult as the mean velocities and wall-normal gradients beyond the boundary layer edge do not exactly go to zero (see also the discussion in \citet{vinuesa2016}). Here, the local boundary layer thickness is defined as the average wall-normal location where the streamwise, wall-normal and Reynolds shear stresses approach an approximately constant value in the freestream. A similar technique using turbulent quantities to identify the boundary layer thickness in open channel flows was also used by \citet{das2022}. 

The streamwise variation of $\delta$ in the present study (see \hyperref[fig:delta_variation]{figure~4}) follows the trend set by alternating pressure gradients discussed previously: a relatively small growth around the leading edge of the cavity due to an APG, followed by a thinning over the windward side of the cavity until the apex due to a FPG and finally a thickening over the leeward side of the cavity due to an APG. This is also in line with trends observed in the studies of \citet{anand2021} and \citet{baskaran1987} (air cavity and solid bump respectively). The TBL here (and in \citet{anand2021}), was found to not separate at the leeward side of the cavity under APG conditions, whereas incipient separation was observed in \citet{baskaran1987} at the leeward side of the solid bump. This is due to the marked difference of the $\delta/t_{max}$ ($\delta/h$) ratio between the three studies, with low values of this ratio typically leading to separation under APG conditions. In order to more closely investigate these alternating pressure gradient effects on the TBL across the cavity, the mean velocity and turbulent stresses profiles are evaluated next, at selected streamwise locations. The profiles have been normalised by local outer units, that is, using the local $\delta$ and $U_{\infty}$ at each streamwise location. As mentioned previously, all profiles are evaluated over the cavity, thus using the local coordinate system $(x', y')$.

\begin{figure}
  \centering 
  \includegraphics[width=0.99\textwidth]{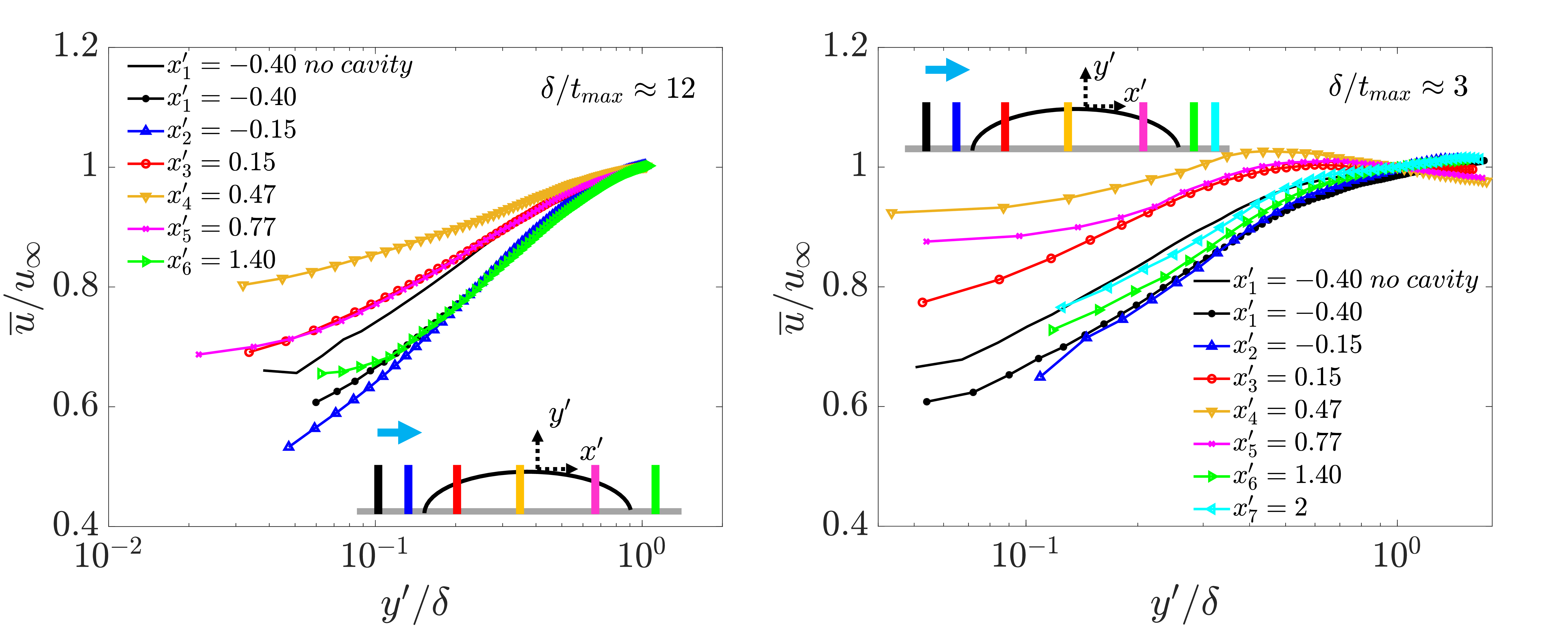}% Images in 100% size
  \caption{Variation of mean streamwise velocity (markers) for $\delta/t_{max}\approx 12$ (left) and $\delta/t_{max}\approx 3$ (right) across different streamwise locations over the cavity, normalised with local outer units. Colours represent different streamwise locations along the cavity, as shown in the inset. The profile at $x'_{1}$ without a cavity is denoted with a black line. Flow is from left to right.}
\label{fig:mean_velocity_variation}
\end{figure}

The mean velocity profiles at different streamwise locations exhibit deviations in the inner and logarithmic regions compared to the baseline case (\hyperref[fig:mean_velocity_variation]{figure~5}, left).  
An upstream influence imposed by the leading edge of the cavity on the incoming boundary layer can be observed at streamwise locations $x_{1}'$ and $x'_{2}$, where an APG causes a deficit in the mean velocity compared to when no cavity is present. This influence has been estimated to last until approximately $0.9\delta$ upstream of the leading edge of the cavity, beyond which the TBL relaxed to its baseline state \citep{anand2021,nikolaidou2024}. 
Over the windward side of the cavity, a switch in the pressure gradient to a favourable one causes the TBL to accelerate, as highlighted by the increase in mean streamwise velocity until the apex (approximately $x'_{4}$).  The profiles display a systematic deviation from log-law behaviour, which was also reported by \citet{narasimha1979} in strongly accelerating flows due to stabilising effects of FPGs. At the leeward side of the cavity ($x'_{5}$), the boundary layer decelerates compared to the mean velocity at $x_{4}'$ due to an APG, however it is still faster than the incoming flow due to the upstream FPG. Overall, the mean velocity profile variation over the cavity follows the trend set by the alternating streamwise pressure gradients, in line with the boundary layer thickness variations discussed previously (see \hyperref[fig:delta_variation]{figure~4}). In an independent experimental campaign in the same setup, but with the air injector located further upstream such that a relatively low $\delta/t_{max} \approx 3$ was achieved \citep{nikolaidou2024}, similar trends in the mean streamwise velocity over the cavity were observed (\hyperref[fig:mean_velocity_variation]{figure~5} right), again with no TBL separation. The recovery of the TBL at $x'_{6}$ back to its state upstream of the cavity (at $x'_{1}$), occurs within one $\delta$ from the trailing edge of the cavity. Further downstream, the return of the TBL to its baseline state (at $x'_{1}$ without a cavity) occurs within a cavity length ($\approx 3\delta$) from the trailing edge of the cavity (\hyperref[fig:mean_velocity_variation]{figure~5}, right).    

\begin{figure}
  \centering 
  \includegraphics[width=0.6\textwidth]{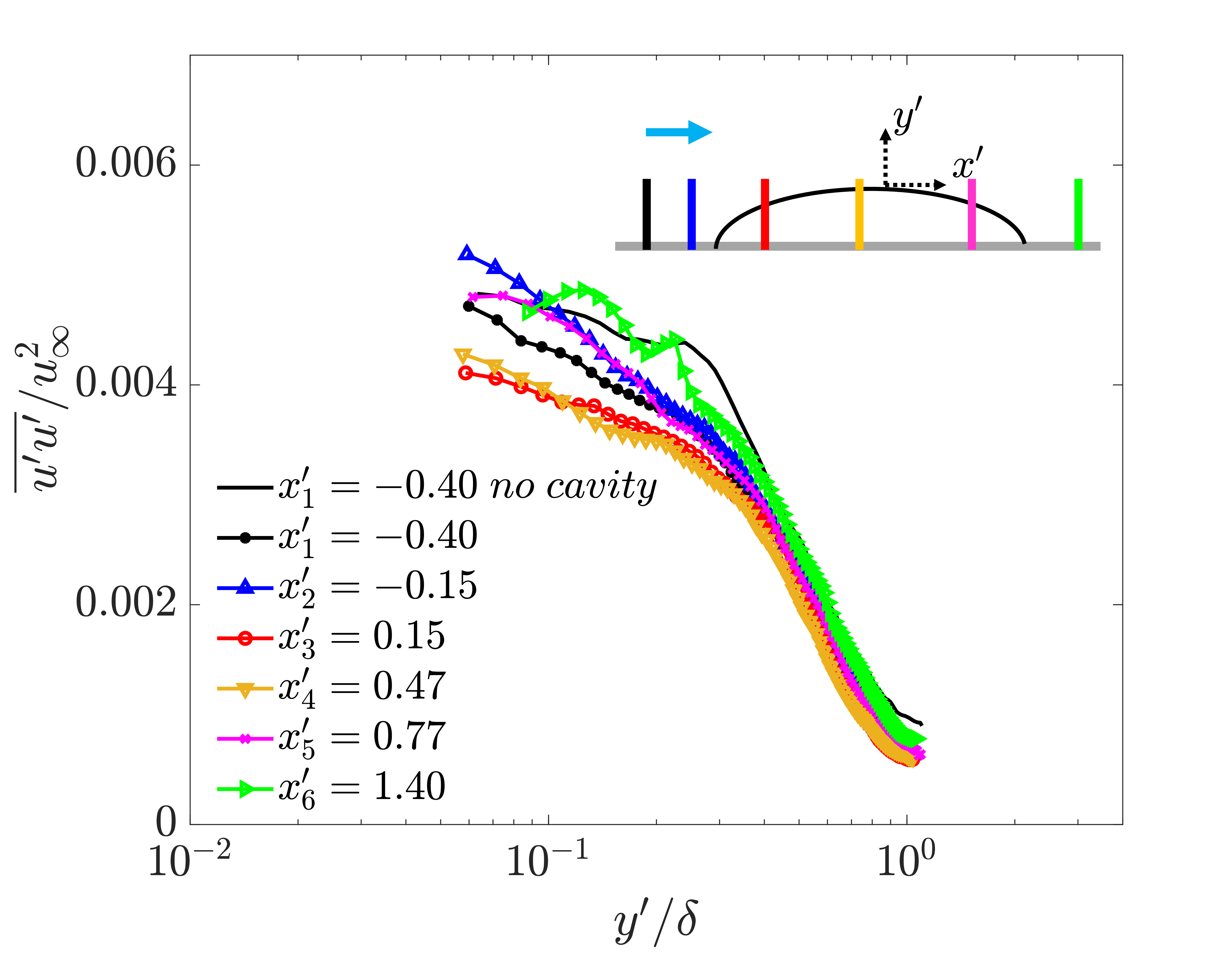}% Images in 100% size
  \caption{Variation of the streamwise normal stress $\overline{u'u'}$ across different streamwise locations over the cavity (markers) normalised with local outer units. Colours represent different streamwise locations along the cavity, as shown in the inset. The profile at $x'_{1}$, when no cavity was present, is denoted with a black line. Flow is from left to right.}
\label{fig:ustress_variation}
\end{figure}

\begin{figure}
  \centerline{\includegraphics[width=0.6\textwidth]{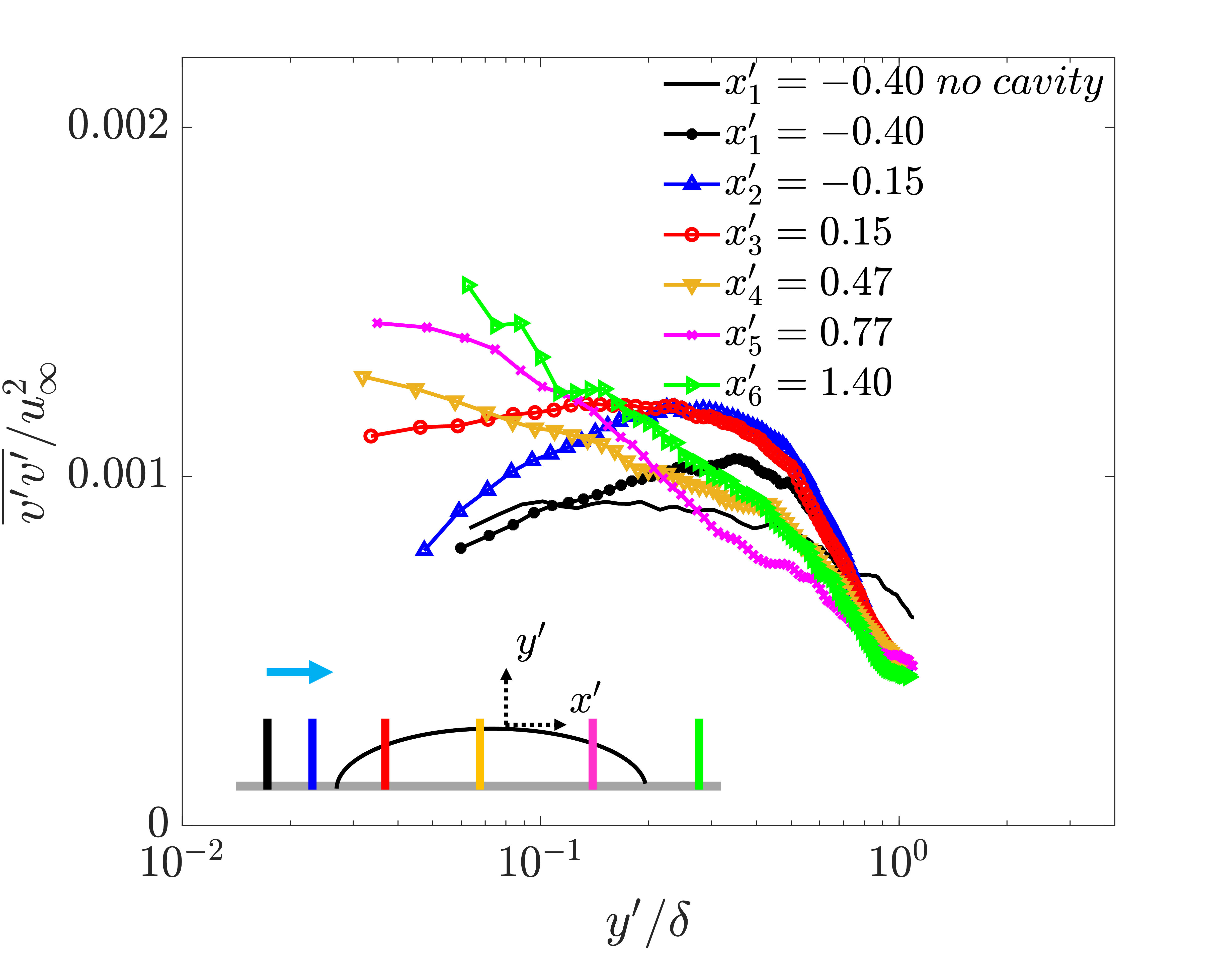}}
  % Images in 100% size
  \caption{Variation of the wall-normal stress $\overline{v'v'}$ across different streamwise positions of the cavity (markers) normalised in local outer units. Colours represent different streamwise locations along the cavity, as shown in the inset. The profile at $x'_{1}$, when no cavity was present, is denoted with a black line. Flow is from left to right.}
\label{fig:vstress_variation}
\end{figure}

\begin{figure}
  \centering 
  \includegraphics[width=0.6\textwidth]{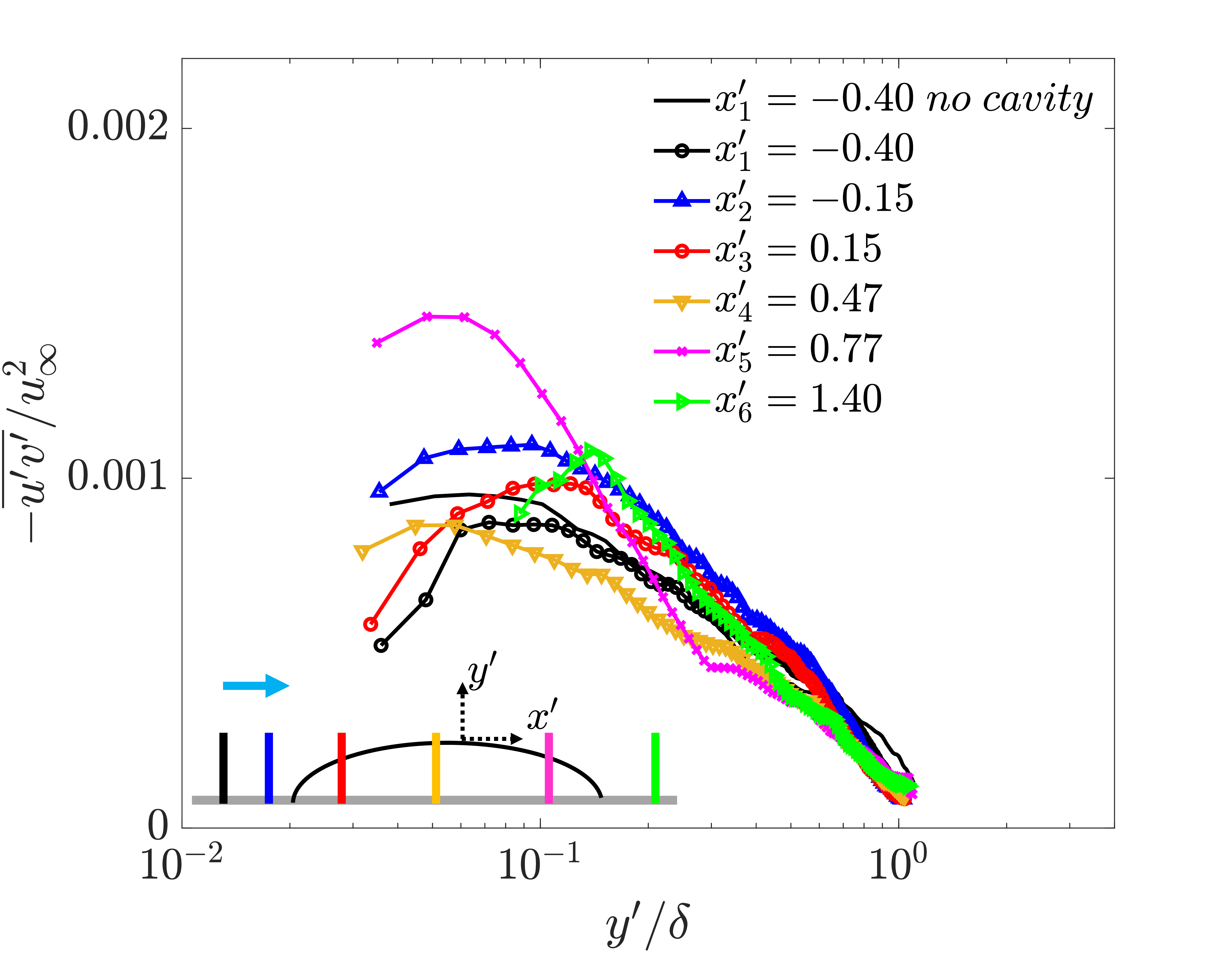}% Images in 100% size
  \caption{Variation of the Reynolds shear stress $-\overline{u'v'}$ across different streamwise positions of the cavity (lines with markers) normalised in local outer units. Colours represent different streamwise locations along the cavity, as shown in the inset. The profile at $x'_{1}$, when no cavity was present, is denoted with a black line. Flow is from left to right.}
\label{fig:rsstress_variation}
\end{figure}

The variation of the streamwise normal ($\overline{u'u'}$, \hyperref[fig:ustress_variation]{figure~6}), wall-normal ($\overline{v'v'}$, \hyperref[fig:vstress_variation]{figure~7}) and Reynolds shear stress profiles (-$\overline{u'v'}$, \hyperref[fig:rsstress_variation]{figure~8}) across the length of the cavity are investigated next. Higher than anticipated reflections from the cavity did not allow for in-depth analysis of the stresses for the prior independent experimental campaign where $\delta/t_{max}\approx 3$ (although qualitatively similar behaviour was observed overall) and as such will not be included in the forthcoming discussion.
The strongest variations as compared with the no cavity case are observed in the $\overline{v'v'}$ and $-\overline{u'v'}$ profiles, and they extend throughout the majority of the TBL height. Variations in $\overline{u'u'}$ on the other hand, are less pronounced (\hyperref[fig:ustress_variation]{figure~6}) and are restricted to $y' < 0.4\delta$, with the rest of the outer region showing a near collapse under outer-scaling. Furthermore, the stress profiles do not exhibit ``knee'' points, which are typical markers for an internal layer triggered within the TBL \citep{webster1996,balin2021,aadhy2023} (see \hyperref[sec: intro]{section~1} ). Internal layers have been reported to be formed in TBLs subjected to relatively strong favorable-adverse pressure gradients (quantified by $K\gtrsim 2\times 10^{-6}$) typically imposed at low $\delta/h$ ratios \citep{aadhy2023}; however, internal layers are not expected to be triggered in the TBL here ($K\approx 0.3\times 10^{-6}$, $\delta/t_{max}= 12$).

As the TBL moves over the windward side of the cavity where it is subjected to a FPG, a suppression in $\overline{u'u'}$ and $-\overline{u'v'}$ profiles throughout the boundary layer is observed as expected \citep{volino2020}. However, $\overline{v'v'}$ exhibits a marked increase below $y\approx 0.3\delta$ throughout the length of the cavity (\hyperref[fig:vstress_variation]{figure~7}), even in the region where an attenuation of stresses is expected due to a FPG (station $x'_{3}$). In addition, the general shape of the $\overline{v'v'}$ profiles over the length of the cavity appears to be largely different in character compared to its state at $x'_1$ and $x'_{2}$.   
The amplification of $\overline{v'v'}$ at $x'_{3}$ and $x'_{4}$ relative to $x'_{2}$ where an attenuation due to a FPG is expected (as seen in $\overline{u'u'}$ and $-\overline{u'v'}$), could be due to the dominance of the vertical momentum induced by the air injection over any pressure gradient effects. A weakening of $\overline{v'v'}$ above $y\approx 0.3\delta$ at $x'_{4}$ and $x'_{5}$ is observed relative to the stations over the windward side of the cavity. Weakening of stresses in the outer region has been attributed to effects of convex streamline curvature \citep{baskaran1987,balin2021} or a combination of convex streamline curvature and FPG effects \citep{uzun2021}. However, the weakening in $\overline{v'v'}$ observed here is not expected to be due to streamline curvature, but is instead solely driven by the effects of FPG. If streamline curvature induced by the shape of cavity was indeed important, we would expect $\overline{u'u'}$ to show a similar decrease at higher distances from the wall ($y' > 0.4\delta$). At a smaller $\delta/t_{max}$ ratio compared to the current study, \citet{anand2021} reported a near collapse above $y\approx 0.4\delta$ in the $\overline{u'u'}$ and $-\overline{u'v'}$ profiles, indicating a minimal impact of streamline curvature. In addition, a similar conclusion was drawn in the solid bump study of \citet{webster1996} with $\delta/h = 1.5$, where only mild variations were observed in the outer region of the stresses. Therefore, we would expect streamline curvature effects induced by the shape of the cavity (or a solid bump) to be even less important at higher $\delta/t_{max}$ ratios. Alternating streamwise pressure gradients and air injection are therefore expected to be the dominant perturbations to the incoming turbulent boundary layer in the current study, the latter being more prominent in the windward side of the cavity.

\FloatBarrier

\begin{figure}[t] 
  \centering 
  \includegraphics[width=0.95\textwidth]{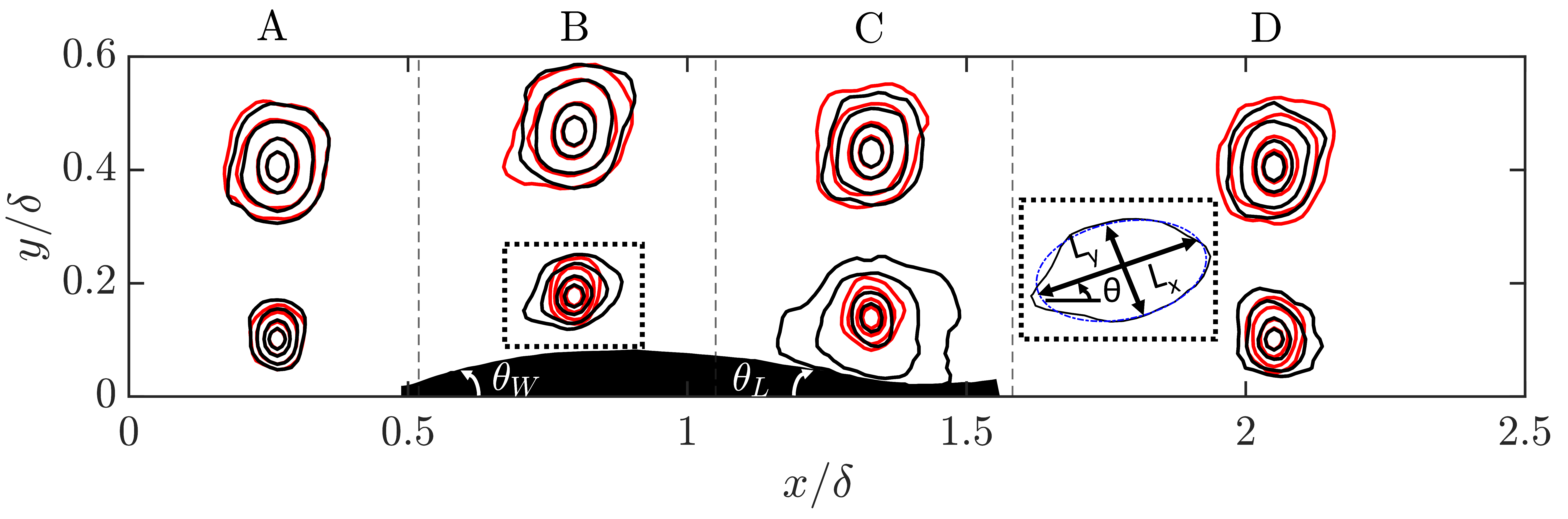}
  
  % Images in 100% size
  \caption{Two-point correlations of wall-normal velocity fluctuations $R_{v'v'}$ at four different streamwise regions ($A$ to $D$) and at two wall-normal locations ($y= 0.1\delta$ and $y= 0.4\delta$), for the cases with (black lines) and without (red lines) the cavity present. Vertical dashed lines demarcate the different streamwise regions where $R_{v'v'}$ is computed. Note that the local $\delta$ is used for normalisation. Contour levels used: [$0.3\;0.4\;0.6\;0.8\;1$]. Inset: Illustration of the elliptical fit (blue dash-dotted line) used for the extent($L_x$ and $L_y$) and orientation ($\theta$) estimates of $R_{v'v'}$ at $0.3$ correlation level (see \hyperref[table:orientation_angle]{table~2}). Approximate inclinations of the cavity: Windward side $\theta_W \approx 14.3^\circ$ and Leeward side $\theta_L \approx -7.7^\circ$. Flow is from left to right.}
\label{fig:two_point_corr}
\end{figure}

Up until this point, only single-point statistics of the TBL have been considered, focusing on magnitude variations across $\delta$ due to the presence of the air cavity; $\overline{v'v'}$ profiles were those impacted the most in both magnitude and overall shape. To further investigate how these variations translate to a change in spatial coherence, two-point correlations of the wall-normal fluctuations ($R_{v'v'}$) were computed (see \hyperref[fig:two_point_corr]{figure~9} and \hyperref[table:orientation_angle]{table~2}), 
at two reference wall-normal locations ($y_{ref}=0.1\delta$ and $0.4\delta$). The extent of $R_{v'v'}$ in the streamwise ($L_x$) and wall-normal direction ($L_y$), was estimated based on an elliptical fit to $R_{v'v'}$ at a correlation level of $0.3$ \citep{wu2010}.   
Qualitatively, with no cavity present, $R_{v'v'}$ was found to be narrow and elongated in the wall-normal direction (red line contour in figure \ref{fig:two_point_corr}) consistent with previous observations of $R_{v'v'}$ over smooth walls \citep{krogstad1994,wu2010}; the spatial extent of $R_{v'v'}$ evaluated at $y_{ref}= 0.15\delta$ and $R_{v'v'} = 0.3$ was found to agree reasonably well with similar estimates in canonical TBLs in literature (Current study: $L_x=0.11\delta$ and $L_y= 0.14\delta$; \citet{wu2010}: $L_x= 0.16\delta$ and $L_y= 0.18\delta$).

In the presence of the cavity, the spatial structure of $R_{v'v'}$ undergoes marked changes as the TBL travels over the cavity. This is more prominently observed at $y_{ref}=0.1\delta$ compared to $y_{ref}=0.4\delta$ (see \hyperref[fig:two_point_corr]{figure~9} and bottom row of \hyperref[table:orientation_angle]{table~2}). The coherence in $R_{v'v'}$ at $y_{ref}=0.4\delta$ remains relatively similar at all streamwise regions examined with no significant changes in the spatial extent of coherence compared to the baseline case (no cavity present). It should be noted here that this similarity in spatial coherence of $R_{v'v'}$ at $y>0.4\delta$ is however coupled with marked changes in magnitude, as shown in the intensity profile evolution of $v'v'$ over the cavity (see \hyperref[fig:vstress_variation]{figure~7}).

\FloatBarrier

\begin{table}
\centering
    \begin{tabular}{ccccccc}
    \toprule
    
     & \textcolor{lightred}{\textbf{A'}} & \textbf{A} & \textbf{B} & \textbf{C} & \textbf{D} & \textbf{$y_{ref}$} \\ 
            & \textcolor{lightred}{\textbf{(no cavity)}} \\
     \toprule 
     \multicolumn{1}{p{1cm}}{\centering $\theta$ \\ $L_x$\\ $L_y$ } 
     & \multicolumn{1}{p{1.8cm}}{\centering $-$ \\ \textcolor{lightred}{$0.18\delta$} \\ \textcolor{lightred}{$0.22\delta$} }
     & \multicolumn{1}{p{1.3cm}}{\centering $-$ \\ $0.18\delta$ \\ $0.22\delta$ }
     & \multicolumn{1}{p{1.3cm}}{\centering $70.1^\circ$ \\ $0.19\delta$\\ $0.23\delta$ }
     & \multicolumn{1}{p{1.3cm}}{\centering $71.2^\circ$ \\ $0.17\delta$\\ $0.21\delta$}
     & \multicolumn{1}{p{3cm}}{\centering $-$ \\ $0.17\delta$ \\ $0.20\delta$} 
     & \multicolumn{1}{p{1.5cm}}{ \centering \; \\ $0.4\delta$ \\ \;  }
     \\\toprule 
      $\theta$  &      $-$           &$-$          & $21.4^\circ$     & -$2.8^\circ$ & $-$          & \\
      $L_x$     &  \textcolor{lightred}{$0.10\delta$}  &$0.10\delta$ & $0.17\delta$     & $0.32\delta$ & $0.15\delta$ & $0.1\delta$ \\ 
      $L_y$     &  \textcolor{lightred}{$0.12\delta$}  &$0.13\delta$ & $0.12\delta$     & $0.23\delta$ & $0.17\delta$ &              \\ 
      \bottomrule 
      \end{tabular}
      \caption{Orientation ($\theta$) and extent ($L_x$ and $L_y$) of $R_{v'v'}$ (at a level of $0.3$, based on \hyperref[fig:two_point_corr]{figure~9}). Note that the streamwise positions chosen (\textbf{A} to \textbf{D}) match those in \hyperref[fig:two_point_corr]{figure~9} for clarity, while the no cavity case at the most upstream position is also added for comparison (denoted with \textcolor{lightred}{\textbf{A'}}).}
      \label{table:orientation_angle}
\end{table}

At $y_{ref}=0.1\delta$, changes to both the aspect ratio and the orientation of $R_{v'v'}$ are observed over the cavity (regions $B$ and $C$ in \hyperref[fig:two_point_corr]{figure~9}), while just upstream of the leading edge (region $A$) they are both relatively similar compared to the baseline case. On average, the spatial structures in regions $B$ and $C$ are found to approximately follow the inclinations imposed by the cavity. In terms of aspect ratio, at region $B$, contours of $R_{v'v'}$ exhibit an increase in the streamwise spatial extent increasing the aspect ratio beyond $1$. This can be attributed to the acceleration imposed by the FPG at the windward side of the cavity: a similar increase in the spatial extent of $R_{v'v'}$ under a FPG was observed in the study of \citet{volino2020}, where the TBL was subjected to changing pressure gradients in a variable-height water tunnel. This anisotropy is further amplified in region $C$, with aspect ratios reaching almost twice those at region $A$, followed by a recovery of $R_{v'v'}$ at region $D$. Considering that the leeward side of the cavity experiences an APG, it is interesting to observe this increased spatial extent of $R_{v'v'}$ at region $C$. This is in contrast to previous studies where a TBL influenced by an APG alone was reported to undergo little to no change in the spatial extent of $R_{v'v'}$ at similar reference wall-normal locations \citep{krogstad1995,gungor2014}. \citet{volino2020} also reported almost no change in $R_{v'v'}$ in an APG region which was preceded by a FPG and a ZPG region.

In an attempt to explain the increased extent of coherence of $R_{v'v'}$ observed at region $C$, we compare the local mean advection ($\mathcal{T}_{adv} \sim  \overline{c}/\overline{u}$) and eddy turnover time ($\mathcal{T}_{eddy} \sim L/u'_{rms}$) scales, $L$ being the spatial extent of $R_{v'v'}$ discussed previously (for comparison, the large eddy turnover time $= \delta/u_{\tau} \sim 5\mathcal{T}_{eddy}$).
The form and response of such structures is expected to be influenced by the competition between $\mathcal{T}_{adv}$ and $\mathcal{T}_{eddy}$. It is expected that for $\mathcal{T}_{adv}>>\mathcal{T}_{eddy}$, structures in the TBL have sufficient time to reach local equilibrium with the local conditions of the cavity interface. As a result, the influence of upstream (history) effects are then expected to be relatively small. On the other hand, at relatively lower $\mathcal{T}_{adv}$, eddies in the TBL do not have sufficient time to reach local equilibrium with the local conditions of the cavity interface. As a result of a lag in their response, upstream influences (and their cumulative effect) are expected to be more important. In the current study we find $\mathcal{T}_{eddy} \sim 2\mathcal{T}_{adv}$. Therefore qualitatively speaking, we expect the structural response of the TBL to be relatively slow to adjust to the local conditions of the cavity interface, bringing possible upstream effects into play. The study of \citet{aadhy2023} involving the application of a sequence of streamwise pressure gradients on a TBL highlighted the importance of the history of pressure gradients influencing the response of the TBL (also see \citet{vinuesa2017}). Therefore, in the present study where the cavity imposes a series of streamwise pressure gradients of alternating signs, the observed increase in $R_{v'v'}$ at region $C$ is expected to be a consequence of a cumulative effect of previous pressure gradients imposed. 
It is important to note that two-point correlations are a statistical measure of the average spatial structure; instantaneously, the spatial coherence is expected to exceed (or decrease from) the estimated extents. Additionally, $R_{v'v'}$ is estimated here in streamwise regions chosen in accordance with the alternating pressure gradients expected around the mean cavity shape (as reflected in the mean and turbulence intensity profiles). The latter, although not significantly changing in geometry, can still impose a pressure gradient switch at streamwise locations which vary instantaneously from the average. It is therefore possible that large-scale structures of the order of half the cavity length or longer (for example, $R_{v'v'}$ at region $C$), are at times subjected to an inhomogeneity in pressure gradients along their spatial extent, thereby making it difficult to distinguish the exact source of the observed deviation from the baseline case. Overall, situations involving a continuously applied APG or FPG or a different sequence of streamwise pressure gradients or even the way the pressure gradients are imposed (for example variable-tunnel heights, ramps, bumps etc.), can alter the statistics and structure of the TBL differently, an area that still requires further exploration.

\begin{figure}
  \centering 
 
  \includegraphics[width=0.97\textwidth]{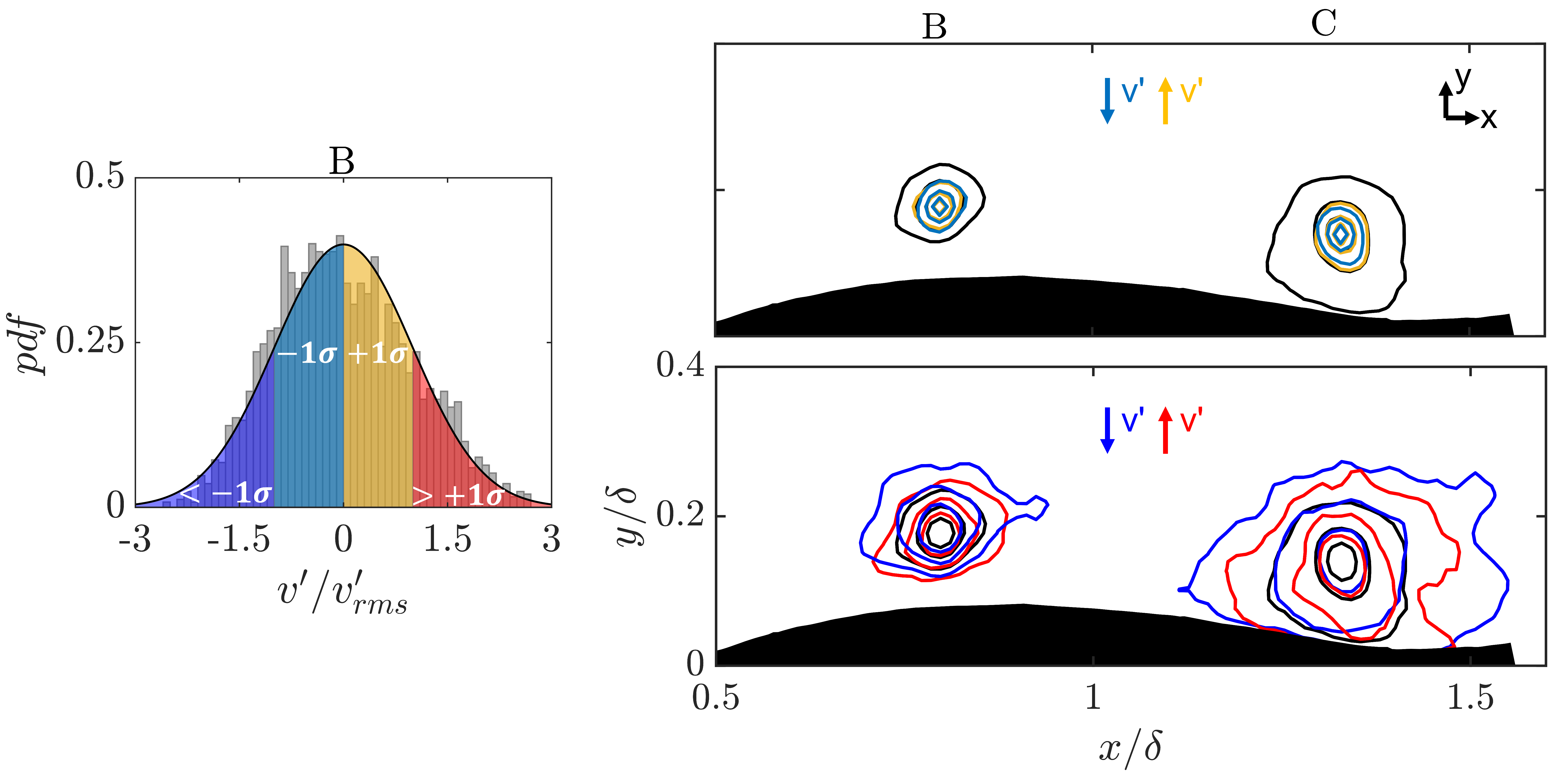}
  \caption{ Left: PDF of the wall-normal velocity fluctuations $v'$ (normalised by the rms $v'_{rms}$) at the windward side of the cavity (region $B$ in \hyperref[fig:two_point_corr]{figure~9}). PDF of $v'$ at the leeward side (region $C$) not shown here for clarity. Differently coloured areas of the PDF correspond to different conditional averages of $R_{v'v'}$: $R|_{v'<v'_{rms}}$ (mustard) and $R|_{v'>-v'_{rms}}$ (light blue); $R|_{v'>v'_{rms}}$ (red) and $R|_{v'<-v'_{rms}}$ (dark blue). Right panel: Contours of the conditioned two-point correlations (at $y_{ref} = 0.1\delta$) at the two streamwise positions and coloured according to the area of the pdf considered (top: $v'<v'_{rms}$ (of either sign); bottom: $v'>v'_{rms}$ (of either sign)). Black lines in the contour represents the unconditioned correlation $R_{v'v'}$ (also shown in the pdf).  Contour levels used: [0.4 0.6 0.8 1]. Flow is from left to right.}
\label{fig:cond_corr}
\end{figure}

In order to further examine the anisotropy observed in the wall-normal velocity coherence around the cavity, $R_{v'v'}$ is conditioned based on the standard deviation of $v'$ ($v'_{rms}$) and the sign of $v'$ ($v'>0$: $R|_{v'<v'_{rms}}$ and $R|_{v'>v'_{rms}}$; $v'<0$: $R|_{v'>-v'_{rms}}$ and $R|_{v'<-v'_{rms}}$, see \hyperref[fig:cond_corr]{figure~10}). The conditional correlations (coloured lines in figure \ref{fig:cond_corr}) are compared to the unconditional $R_{v'v'}$ (black lines in \hyperref[fig:two_point_corr]{figure~9} and \hyperref[fig:cond_corr]{figure~10}). The spatial coherence and anisotropy observed in the unconditional $R_{v'v'}$ are clearly due to high intensity events. For fluctuations with intensities $<v'_{rms}$ (of either sign), spatial coherence is limited and more isotropic compared to the global behaviour, while those with intensities $>v'_{rms}$ (of either sign), a significant elongation in $x$ is observed, leading to the observed anisotropy in the unconditional correlations. At the leeward side of the cavity, this high intensity, anisotropic behaviour is also seen to be prominent for $v'<0$, thus for structures moving towards the cavity interface (see \hyperref[table:cond_corr]{table~3} for an estimate). 
These relatively large coherent structures (with $v'<0$) near the trailing edge of the cavity (where the cavity is observed to breakup and shed), can play an important role in the cavity's form and stability: regions of downwards moving liquid (directed towards the cavity) can lead to pinch-offs at the cavity's trailing edge, promoting its breakup and resulting in a shortened length. Along with this mechanism, other factors like the presence of re-entrant jets and small-scale wave instabilities at the cavity interface proposed in literature (see for example \citet{makiharju2013,zver2014}), can shorten the cavity length. This shortening can have negative implications in ship air lubrication applications, where a minimal contact area between water and solid surface is sought, that is, achieving and maintaining the maximum possible cavity lengths. To establish the extent of such interactions however, simultaneous time-resolved measurements of the instantaneous cavity shape and stability and TBL structuring would be required, which is out of the scope of the present study.

\begin{table}
\centering
\begin{tabular}{*5c}
\toprule
 &  \multicolumn{2}{c}{Windward side} & \multicolumn{2}{c}{Leeward side}\\
\midrule
{}   & $R|_{v'<-v'_{rms}}$   & $R|_{v'>v'_{rms}}$    & $R|_{v'<-v'_{rms}}$   & $R|_{v'>v'_{rms}}$\\
$L_x$   &  $0.21\delta$ & $0.17\delta$   & $0.41\delta$  & $0.29\delta$\\
$L_y$   &  $0.13\delta$ & $0.12\delta$   & $0.25\delta$  & $0.23\delta$\\
\bottomrule
\end{tabular}
\caption{Length scale ($L_x$ and $L_y$) estimates of conditioned $R_{v'v'}$: $R|_{v'>v'_{rms}}$ (red) and $R|_{v'<-v'_{rms}}$) (dark blue) contours in \hyperref[fig:cond_corr]{figure~10} (right panel bottom) measured at $0.4$ correlation level using an elliptical fit (see \hyperref[fig:two_point_corr]{figure~9}). $L_x$ and $L_y$ are lengths along the streamwise and wall-normal directions respectively. }
\label{table:cond_corr}
\end{table}

\section{Conclusions}
\label{sec:conclusion}

The turbulent boundary layer (TBL) development over an air cavity was experimentally investigated using planar particle image velocimetry. These type of flows are typically encountered in the application of ship air lubrication where the air cavity dynamics and morphology depend on the incoming flow conditions and vice versa. The present study provides for the first time an experimental insight into a subset of such flows: the response and development of a TBL to an air cavity formed at a specific incoming flow condition and air flow rate. The flow over the air cavity shares similarities to flow over solid bumps in literature \citep{baskaran1987,webster1996}. However, unlike solid bumps which have a fixed maximum thickness and length, the air cavity has a relatively variable geometry inherent to its dynamic nature. In addition, the presence of multiple pressure gradients and curvatures including a free-slip boundary, makes the behaviour of the TBL significantly more complex. Given the important role played by the geometry of solid bumps in the development of the incoming flow observed previously in literature, the instantaneous geometries of the cavity were identified using a detection technique based on the thresholding of correlation values. The detection revealed a well-defined, asymmetric bump-like geometry at the spanwise-uniform region of the cavity also supported by visual observations. Statistics of the cavity geometry revealed the amplitude of the cavity thickness fluctuations across the cavity length to be approximately constant to within experimental uncertainty, indicating that there was no significant deviation of the cavity from a bump-like geometry. The ratio of the incoming boundary layer thickness to the maximum thickness of the cavity was found to be approximately $\delta/t_{max} = 12$, which is considerably higher than ratios encountered in the solid bump studies of \citet{baskaran1987} and \citet{webster1996} ($\delta/h = 0.25$ and $\delta/h = 1.5$ respectively, $h$ being the height of the bump). Consequently, the TBL did not undergo separation at the leeward side of the cavity.

The boundary layer thickness ($\delta$) and the mean streamwise velocity profiles, were found to follow the trend set by alternating streamwise pressure gradients, consistent with previous works on flow over solid bumps: from an adverse pressure gradient (APG) around the cavity leading edge, followed by a favourable pressure gradient over the windward side of the cavity and finally an APG over the leeward side of the cavity. Trends set by turbulence stresses (normalised in outer-scales) particularly, $\overline{u'u'}$ and $-\overline{u'v'}$, exhibited expected behaviours set by alternating streamwise pressure gradients: a weakening of stresses in the FPG region and an amplification of stresses in the APG region, mostly occurring in the lower part of the outer layer. The behaviour of $\overline{v'v'}$, however, appeared to be largely different in character over the cavity: it was observed to amplify below $y\approx 0.3\delta$ throughout the length of the cavity, even in the region where an attenuation of stress was expected due to a FPG. The amplification of $\overline{v'v'}$ was attributed to the dominance of the vertical momentum introduced by the air injection over any pressure gradient effects, particularly over the windward side of the cavity. Streamline curvature on the other hand, was concluded to be marginal, based on previous observations. The recovery of the TBL to its state upstream of the cavity is rapid, occurring within a boundary layer thickness. However, effects of the APG imposed by the leading edge of the cavity remain present in the TBL in comparison to its baseline state (no cavity). 

Two-point correlations of the wall-normal velocity fluctuations ($R_{v'v'}$) at $y_{ref}=0.4\delta$ exhibited similar extent of coherence at all streamwise regions over the cavity, with no significant changes from the baseline case (no cavity). A local anisotropy and changes to the orientation of $R_{v'v'}$ (following approximately the inclinations of the cavity) were observed at $y_{ref}=0.1\delta$ though,  more pronounced at the leeward side: the aspect ratio nearly doubled and the length reached the order of half the cavity length. The increased coherent extent of $R_{v'v'}$ in a region subjected to an APG, in contrast to studies where a TBL was subjected to an APG alone, was attributed to the cumulative effect of multiple streamwise pressure gradients, resulting from a relatively slow response of the TBL to the local conditions of the cavity interface. The anisotropy in $R_{v'v'}$ due to the increased spatial coherence in $x$ was further shown to be be a result of high amplitude events ($v'>v'_{rms}$) and more pronounced for $v'<0$ (moving towards the cavity) , especially at the leeward side. These relatively large coherent regions of liquid directed towards the cavity can play an important role in the formation of the latter, an aspect that is crucial in the application of ship air lubrication.

To sum up, the present study provides an insight into the physics of one aspect of the liquid-gas two-way coupling, namely, the response and development of a TBL in the presence of the air cavity; the complexity and lack of relevant studies on the topic, highlight the need for future work. Further rigorous investigations with time-resolved measurements of the instantaneous cavity shape and the surrounding velocity field are necessary to study their interaction and isolate the different ingredients at play, such as alternating pressure gradients, surface curvature and the effects of air injection and a free-slip boundary condition, the latter aspects setting this family of flows apart from solid bump flows. Results and conclusions from this study are expected to not only provide insights regarding cavity formation, but also establish a baseline reference for geometries involving flow over gaseous layers while serving as a testbed for future studies examining variable inflow conditions and external perturbations.

\clearpage

\begin{Backmatter}

\paragraph{Acknowledgements}
The authors would like to gratefully acknowledge the help of Edwin Overmars and Jasper Ruijgrok in setting up the experiments at the Laboratory of Aero and Hydrodynamics at TU Delft. 

\paragraph{Funding Statement}
This research received no specific grant from any funding agency, commercial or not-for-profit sectors. 

\paragraph{Declaration of Interests}
The authors declare no conflict of interest.

\paragraph{Author Contributions}
\textit{Abhirath Anand}: Investigation (lead), Conceptualization, Methodology, Visualization, Validation, Formal analysis, Writing -- original draft.  
 \textit{Lina Nikolaidou}: Investigation (support), Conceptualization, Methodology, Writing -- review \& editing.  \textit{Christian Poelma}: Conceptualization, Methodology, Supervision, Writing -- review \& editing.  \textit{Angeliki Laskari}: Conceptualization, Methodology, Supervision, Writing -- review \& editing.

\paragraph{Data Availability Statement}
Data will be made available on request upon publication.  

\paragraph{Ethical Standards}
The research meets all ethical guidelines, including
adherence to the legal requirements of the study country.

\paragraph{Supplementary Material}
Supplementary movie not available on arXiv.

\bibliographystyle{jfm}
\bibliography{references}

\end{Backmatter}

\end{document}